\documentclass{ws-procs11x85}

\makeindex
\begin{document}


\title{MIXINGS, LIFETIMES, SPECTROSCOPY AND 
PRODUCTION \\ OF HEAVY FLAVOR AT THE TEVATRON}

\author{K.~T.~PITTS}

\address{University of Illinois, Department of Physics,
1110 West Green St., Urbana, IL 61801, USA 
\\E-mail: kpitts@uiuc.edu}


\twocolumn[\maketitle\abstract{The Fermilab Tevatron 
offers unique opportunities to perform measurements
of the heavier $B$ hadrons that are not accessible at
the $\Upsilon(4S)$ resonance.  In this summary, we describe
some recent heavy flavor results from the D\O\ and CDF 
collaborations and discuss prospects for future measurements.}]

\baselineskip=13.07pt

\section{Introduction}

In the 1980's and 1990's, the complementary $b$ physics 
programs of CLEO, ARGUS,
the LEP experiments, SLD, D\O\  and CDF  began to make 
significant contributions to our understanding of the production and decay
of $B$ hadrons.  With the successful turn-on of the BaBar and Belle
experiments in the last few years, many experimental measurements in
the $b$ sector have achieved impressive precision.

A number of questions remain, and it will again take an effort of
complementary measurements to make further progress on our understanding
of the $b$ system.  The Fermilab Tevatron, along with the 
upgraded CDF and D\O \  detectors, offers a unique opportunity to study
heavy flavor production and decay.  In many cases, the measurements that
can be performed at the Tevatron are complementary to those performed
at the $e^+e^-$ $B$-factories.

In this document, we summarize some of the recent experimental 
progress in the measurements of $B$ lifetimes, spectroscopy, mixing,
and heavy flavor production at the Tevatron.  Since the details of 
many of 
these analyses are presented in other publications, we will attempt 
to include 
some background information in this summary that the reader might
not find elsewhere.

\section{Overview:  $B$ Physics at the Tevatron}

In $p\overline{p}$ collisions at $\sqrt{s}=1.96\, \rm TeV$,
the $b\overline{b}$ cross section is large
${\cal O}(50 \mu b)$, yet it is only about $1/1000^{\rm th}$ 
the total inelastic $p\overline{p}$ cross section.  At a
typical Tevatron instantaneous luminosity of
${\cal L} = 4\times 10^{31}\, \rm cm^{-2}s^{-1}$, 
we have a $b\overline{b}$ production rate of $\sim\! 2~{\rm kHz}$
compared to an inelastic scattering rate of 
$\sim\! 2~{\rm MHz}$.

\begin{table}
\caption{$B$ mesons and baryons.  This is an incomplete 
list, as there are excited states of the mesons and baryons
(\it e.g. \rm ${B^*}^0$).  Also, a large number of $b$-baryon
states (\it e.g. \rm $\Sigma^-_b = | ddb >$)
and bottomonium states (\it e.g. \rm $\eta_b$, $\chi_b$)
are 
not listed.}
\label{ta:b}
\begin{center}
\begin{tabular}{lll}
Name & $\overline{b}$ hadron & $b$ hadron  \\
\hline 
charged $B$ meson &
$B^+ =|\overline{b}u>$  &
       $B^- = |b\overline{u}>$ \\
neutral $B$ meson & 
$B^0 = |\overline{b}d>$ & $\overline{B^0} = |b\overline{d}> $\\
$B_s$ ($B$-sub-$s$) meson  & 
 $B^0_s = |\overline{b}s>$ &  $\overline{B^0_s}=|b\overline{s}>$\\
$B_c$ ($B$-sub-$c$) meson &  $B^+_c = |\overline{b}c>$ &  
$B^-_c = |b\overline{c}>$ \\  
$\Lambda_b$ (Lambda-$b$) &
  $\overline{\Lambda_b} = |\overline{u}
\overline{d}\overline{b}> $
& $\Lambda_b = |udb> $ \\
$\Upsilon$ (Upsilon) & \multicolumn{2}{c}{$\Upsilon = |\overline{b}b>$}  \\
\hline
\end{tabular}
\end{center}
\end{table}

The $b\overline{b}$
quarks are produced by the strong interaction, which preserves
``bottomness'', therefore they are always produced in pairs.\footnote{It is
possible to produce $b$'s singly through weak decays such as
 $W^-\rightarrow \overline{c}b$ and $W^-\rightarrow\overline{t}b$. The 
cross sections for these processes are several orders of magnitude
below the cross section for direct $b\overline{b}$ production by
the strong interaction.}  
Unlike $e^+e^-$ collisions
on the $\Upsilon(4S)$ resonance, the high energy collisions 
access all angular momentum states, so the $b$ and $\overline{b}$
are produced incoherently.  As a consequence, lifetime, mixing
and {\it CP}-violation measurements can be performed by reconstructing
a single $B$ hadron in an event.  
The produced
$b$ quarks can fragment into all possible species of $B$ hadrons, 
including $B_s$, $B_c$ and $b$-baryons.     Table~\ref{ta:b}
lists the most common species of $B$ hadrons,  all of which
 are accessible at the Tevatron.

The transverse momentum ($p_T$) spectrum for the produced $B$ hadrons is 
a steeply falling distribution, which means that most of the $B$ hadrons 
have very
low transverse momentum.  For example,
a fully reconstructed sample of $B\rightarrow J/\psi K$
decays has an average $B$ meson $p_T$ of about $10\, {\rm  GeV}/c$.
As a consequence, the tracks from these decays 
are typically quite soft, often having $p_T < 1\, {\rm GeV}/c$.  One
of the experimental limitations 
in reconstructing these modes is the ability to 
find charged tracks at very low momentum.
$B$ hadrons with low transverse momentum
do not necessarily 
have low total momentum.  Quite frequently,
the $B$ mesons have very large longitudinal momentum 
(the momentum 
component along the beam axis.)  These $B$ hadrons are boosted along
the beam axis and are consequently outside the acceptance of the detector.

To reconstruct the $B$ hadrons that do fall within the detector
acceptance, the experiments need excellent 
tracking that extends down to low transverse momentum, 
excellent vertex detection to identify the long-lived hadrons
containing heavy flavor, and high-rate trigger and data acquisition
systems to handle the high rates associated with this physics.
In the following section, we outline some of the relevant aspects
of the Tevatron detectors.

\section{The CDF and D\O\  Detectors}

The CDF and D\O\ detectors are both axially symmetric
detectors that cover about $98\% $ of the full $4\pi$
solid angle around the proton-antiproton interaction
point.
For Tevatron Run~II, 
both experiments have axial solenoidal magnetic  fields, central
tracking, and silicon microvertex detectors.
Additional details about the experiments can be found
elsewhere\rlap.\,\cite{d0,cdf}  The strengths of the detectors are somewhat 
complementary to one another and are discussed briefly below.

\subsection{D\O}
The D\O\ tracking volume features a $2\, \rm T$ solenoid
magnet surrounding  a scintillating fiber central
tracker that covers the region  $|\eta| \le 1.7$, where $\eta$ is
the pseudorapidity, $\eta = -ln(tan(\theta/2))$, and $\theta$
the polar angle measured from the beamline.  The D\O\ silicon
detector has a barrel geometry interspersed with disk 
detectors which extends  the forward tracking to $|\eta| \le 3$. 
In addition, the D\O \ muon system covers $|\eta| \le 2$ and
the  uranium/liquid-argon
calorimeter  has very good energy resolution
for electron, photon and hadronic jet energy measurements.

\subsection{CDF}
The CDF detector features a $1.4\, \rm T$ solenoid
surrounding a silicon microvertex detector and
gas-wire drift chamber.  The CDF spectrometer has 
excellent mass resolution.  These properties, combined
with muon detectors and calorimeters, allow
for excellent muon and electron identification, as
well as precise tracking and vertex detection
for $B$ physics.

\subsection{Triggering}

Both experiments exploit heavy flavor decays which
have leptons in the final state.  Identification of
dimuon events down to very low momentum is possible,
allowing for efficient $J/\psi \rightarrow \mu^+\mu^-$
triggers.  As a consequence, both experiments are able
to trigger upon the $J/\psi$ decay, and then fully 
reconstruct decay modes such as $B^0_s\rightarrow J/\psi \phi$,
with $\phi\rightarrow K^+K^-$.  Triggering on dielectrons
to isolate $J/\psi \rightarrow e^+e^-$ decays 
is also possible, although at low momentum the backgrounds
become more problematic.  

CDF has implemented a 
$J/\psi\rightarrow e^+e^-$ trigger requiring each electron
have $p_T>2\, \rm GeV/{\it c}$.  Because the triggering and
selection cuts required to reduce background are more stringent
than they are for the dimuon mode, the yield for the
$J/\psi\rightarrow e^+e^-$ mode is about $1/10^{\rm th}$ the yield
for the dimuon mode.  However, since selection criteria
isolate a dielectron mode is
at higher momentum then the dimuon mode, 
the $B$ purity in the $J/\psi\rightarrow e^+e^-$ channel is 
higher.   
The analyses shown here use only the dimuon mode, although
future analyses will supplement the signal sample with the
dielectron mode.

Both experiments also have inclusive lepton triggers designed
to accept semileptonic $B\rightarrow \ell \nu_\ell X$ decays.
D\O\ has an inclusive muon trigger with excellent acceptance,
allowing them to accumulate very large samples of semileptonic
decays.  The CDF semileptonic triggers require an additional
displaced track associated with the lepton,  providing 
cleaner samples  with a smaller yields.

\begin{figure}
\center
\psfig{figure=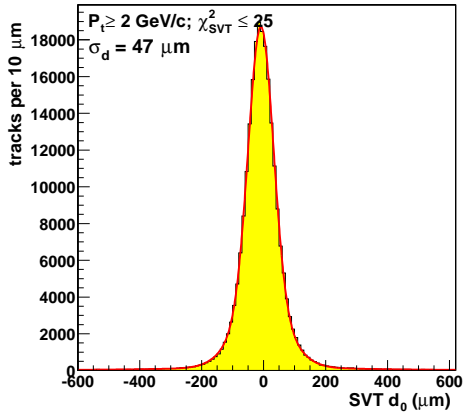,width=6.00truecm}
\caption{Resolution for the CDF Silicon Vertex Tracker (SVT.)
The SVT performs axial tracking in the trigger 
using the silicon microvertex detector.  This plot shows the
impact parameter resolution for a sample of tracks.  The
resolution is $47\,\mu{\rm m}$.  
The typical heavy flavor trigger 
requires two tracks with impact parameter greater than
$120\, \mu{\rm m}$, which has a rejection factor that is 250 times greater
than simply demanding charged tracks.}
\label{fig:svt}
\end{figure}

New to the CDF detector is the ability to select events
based upon track impact parameter.  The CDF Silicon Vertex
Tracker (SVT) operates as part of the Level~2 trigger 
system.  Tracks identified by the eXtremely Fast Tracker (XFT)
are passed to the SVT, which appends silicon hits to the
tracks to measure the impact parameter of each track.
With the high trigger rate, it is very challenging to 
extract the data from the silicon detector and perform
pattern recognition quickly.  The SVT takes on average $25\, \mu\rm s$
per event to extract the data from the silicon, perform silicon
clustering and track fitting.
 As shown in Fig.~\ref{fig:svt}, the impact 
parameter resolution for tracks with $p_T>2\, {\rm GeV}/c$ is
$47\, \mu\rm m$, which is a combination of the primary beam
spot size ($30\, \mu m$) and the resolution of the
device ($35\, \mu m$.)  The CDF SVT has already shown that
it will provide a number of new modes in both bottom and charm
physics that were previously not accessible.  D\O\ is currently
commissioning a displaced track trigger as well.

\section{Heavy Flavor Yields}

As mentioned in the previous section, the $b\overline{b}$
production cross section is very large at the Tevatron.
Although triggering and reconstruction is very challenging,
large samples can indeed be acquired.  Table~\ref{ta:yields} shows
approximate yields for several different modes at the 
Tevatron.\footnote{Throughout this note, we will write charge
specific decay modes for clarity.  All analyses presented here
include the charge-conjugate modes.}   As of this writing, each experiment has logged
approximately ${\cal L}= 200 \, \rm pb^{-1}$ of integrated
luminosity.  The data sample is expected to more than 
double during the 2004 running period.

\begin{table}
\caption{Tevatron yields for various fully- and 
partially-reconstructed heavy flavor modes.
The column labeled ``trigger'' lists the primary signature
utilized to select the events.  The trigger type $d_0$ is
referring to the impact parameter described in the text.
This list is not meant to be exhaustive, only to give the 
reader
a feel for the sample sizes for heavy flavor analyses at the
Tevatron.  A number of $J/\psi$, semileptonic and all-hadronic
decay modes are not included in this table.}
\label{ta:yields}
\begin{center}
\begin{tabular}{cccc}
\hline
Hadron & Decay mode & Trigger & Yield per pb$^{-1}$ \\
\hline

$D^0$ & $K^-\pi^+ $ & $d_0$  & 6000 \\
$D^+$ & $K^-\pi^+\pi^+$ & $d_0$ & 5000 \\
$B^-$ & $D^0\pi^- $ & $d_0$ & 16 \\
$B^0_s$ & $ D^-_s \pi^+ $ & $d_0$ & 1 \\
$B^0_s $ & $ K^+K^- $ & $d_0$ & 1.5 \\ 
$J/\psi $ & $\mu^+\mu^-$ & dimuon & 7000 \\
$B^+$ & $J/\psi K^+$ &  dimuon & 11 \\
$B_s$ & $J/\psi \phi$ & dimuon  & 1 \\
$\Lambda_b$ & $J/\psi \Lambda $ & dimuon & 0.7 \\
$B$ & $D\ell\nu $ & single lepton & 400 \\
$\Lambda_b$ & $\Lambda_c \ell \nu $ & single lepton & 10 \\ 
\hline
\end{tabular}
\end{center}
\end{table}

\section{Prompt Charm Cross Section}

Previously published measurements of the $b$ production
cross section at the Tevatron have consistently been 
significantly higher than the Next-to-Leading Order 
QCD predictions.  Although there has been theoretical
activity in this arena and the level of the discrepancy
has been reduced, it is not yet clear that the entire
scope of the problem is fully understood.  Both experiments
will again measure the $b$ and $b\overline{b}$ cross 
sections at higher center-of-mass energy.

\begin{figure}
\center
\psfig{figure=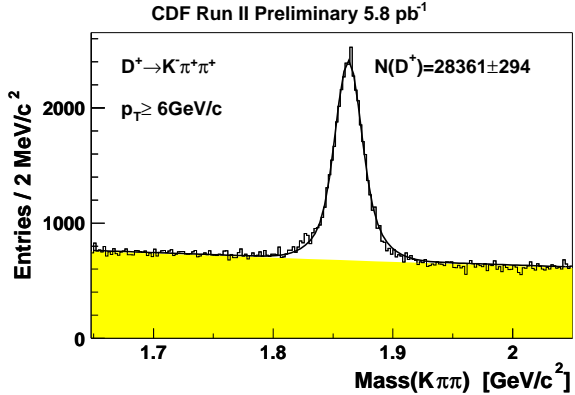,width=8.0truecm}
\caption{Yield for $D^+\rightarrow K^-\pi^+\pi^+$ used in the
charm cross section analysis.  Yields of the size shown in the
plot are routinely acquired in a single week of data taking.  
This sample was acquired using the CDF SVT trigger. About
80\% of the signal in this sample arises from direct charm 
production.}
\label{fig:charm}
\end{figure}

To further shed light on this problem, CDF has recently
presented a measurement of the charm production cross
section\rlap.\,\cite{charmxsec}
Using the secondary vertex trigger, CDF has been able to
reconstruct very large samples of charm decays.  Figure~\ref{fig:charm}
shows a fully 
reconstructed $D^+\rightarrow K^-\pi^+\pi^+$ signal using
$5.8\, \rm pb^{-1}$ of data from early in the run.  
In the full data sample available at
the time of this Symposium (${\cal L}\sim \! 200\, \rm pb^{-1}$),
CDF has a $D^0$ sample exceeding 2 million events.  As this
sample grows, competitive searches for {\it CP}-violation in the 
charm sector and $D^0/\overline{D^0}$ mixing are anticipated.

Since the events are accepted based upon daughter tracks
with large impact parameter, it is clear that the sample 
of reconstructed charm decays contains charm from bottom
($p\overline{p}\rightarrow b\overline{b}X$, with 
$b\rightarrow c \rightarrow D$) in addition to prompt 
charm production ($p\overline{p}\rightarrow c\overline{c}X$, with 
$c\rightarrow D$.)  To extract the charm meson cross section,
it is necessary to extract the fraction of $D$ mesons that
are coming from prompt charm production.  This is done by
measuring the impact parameter of the charm meson.  If it 
arises from direct $c\overline{c}$ production, the charm 
meson will have a small impact parameter pointing back
to the point of production, which was the collision vertex.
If the charm meson 
arises from $b$ decay, it will typically 
not extrapolate back to the primary vertex.

Using this technique, along with a sample of $K^0_S\rightarrow \pi^+ \pi^-$
decays for calibration, it is determined that 80-90\% (depending 
upon the mode) of the 
charm mesons arise from direct charm production.  The shorter
charm lifetime is more than compensated by the copious charm
production in the high energy collisions.

\begin{figure}
\center
\psfig{figure=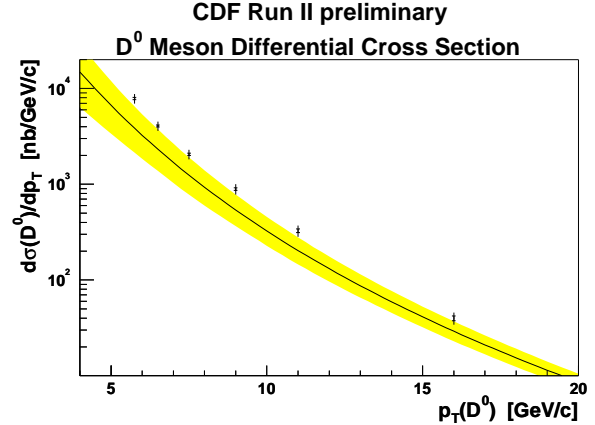,width=8.50truecm}
\caption[]{The measured differential cross section for prompt 
$p\overline{p}\rightarrow
D^0 X$, with $|y(D^0)|<1$.  The $D^0$ hadrons arising from $B$ decays
have been removed.  The NLO calculation is from Cacciari and Nason\rlap.\,\cite{nlo}
The results for the other charm modes are summarized in Table~\ref{ta:charm}.
}
\label{fig:cxsec}
\end{figure}

The full analysis includes measurements of the differential
cross sections for prompt $D^0$, $D^+$, $D^{*+}$ and $D^+_s$ meson
production.  The integrated cross section 
results of this study are summarized in Table~\ref{ta:charm}.
Figure~\ref{fig:cxsec} shows the comparison between data and
the NLO calculation for the differential 
$D^0$ cross section\rlap.\,\cite{nlo}  The trend seen
in this figure is the same for the other $D$ species.  The 
prediction seems to follow the measured cross section in 
shape, but the absolute cross section is low compared to the
measured results.  This difference in magnitude between the
measured and predicted charm meson cross section is similar to
the difference in data and theory seen in the $B$ meson cross sections.

As an interesting aside, we can also compare the measured
 $B$ and $D$ cross sections at the Tevatron\rlap.\,\cite{bxsec}  Looking at the
charged mesons, the measured cross sections are:
\begin{displaymath}
\sigma(D^+, p_T\ge 6~ {\rm GeV}/c, |y|\le 1)= 4.3\pm 0.7~\mu{\rm b}
\end{displaymath}
\vspace*{-0.25truein}
\begin{displaymath}
\sigma(B^+, p_T\ge 6~ {\rm GeV}/c, |y|\le 1)= 3.6\pm 0.6~\mu{\rm b}. 
\end{displaymath}
For this momentum range, the $D^+$ cross section is only 
$20\%$ larger than the $B^+$ cross section.  At very high 
transverse momentum (corresponding to high $Q^2$), we
would expect that the mass difference between the bottom and
charm quarks to be a small effect, yielding similar production
cross sections.  However these results show that even at
lower $p_T$, the mass effects are not that significant.

\begin{table}
\caption{CDF measurement of the direct charm cross section.  The results
are for $D$ mesons with $|y|<1$.}
\label{ta:charm}
\begin{center}
\begin{tabular}{ccc}
Meson & Momentum range & Measured cross section(pb)  \\
\hline 
$D^0$ & $p_T>5.5 \, \rm GeV/{\it c}$  & $13.3\pm 0.2\pm 1.5 $\\
$D^{*+}$ &$p_T>6.0 \, \rm GeV/{\it c}$ &  $5.2\pm 0.1\pm 0.8 $ \\
$D^+ $ &$p_T>6.0 \, \rm GeV/{\it c}$ &  $4.3\pm 0.1\pm 0.7 $\\
$D^+_s$ &$p_T>8.0 \, \rm GeV/{\it c}$ &  $0.75\pm 0.05\pm 0.22 $\\
\hline
\end{tabular}
\end{center}
\end{table}

\section{$B$ Lifetimes} 

\label{se:lifetimes}

In the spectator model for meson decays, where
the light quark does not participate in the decay,
all $B$ lifetimes are equal, since the lifetime is 
exclusively determined by the lifetime of the $b$ quark.  In
reality, non-spectator effects such as interference
modify this expectation.  The Heavy Quark Expansion
(HQE) predicts the lifetime hierarchy for the $B$ 
hadrons as:

\begin{equation}
\tau (B^+) > \tau (B^0) \simeq \tau (B_s) > \tau (\Lambda_b) >>
\tau (B_c),
\end{equation}
where the $B_c$ meson is expected to have the shortest
lifetime because both the $b$ and
the $c$ quarks are able to decay by the weak interaction.

The lifetimes of the light mesons, $B^0$ and $B^+$,
are measured with a precision that is better than
$1\%$.  This impressive level of precision is 
dominated by the measurements of the Belle and Babar
experiments\rlap.\,\cite{taubabar}  The $B_s$ and $b$-baryon 
lifetimes have been measured
by the LEP, SLD and CDF Run I experiments.  One 
interesting puzzle that persists from those measurements
is that the  $\Lambda_b$ lifetime is significantly
lower than expectation\rlap.\,\cite{hfag}

To measure the lifetime of a particle, the experiments 
utilize their precision silicon tracking to measure the
flight distance of the hadron before it decays.  At the
Tevatron, this is done in the plane transverse to the
beamline, and the
two-dimensional flight distance is denoted as $L_{xy}$.
Since the particle is moving at a high velocity in the
lab frame, the decay time measured in the laboratory is
dilated relative to the proper-decay time, which is the 
decay time of the particle in its rest frame.  To extract
the proper decay time, we must correct for the time 
dilation factor:

\begin{equation}
ct_{decay} = {L_{xy}\over{(\beta \gamma)_T}},
\end{equation}
where $t_{decay}$ is the proper decay time in the rest
frame of the particle, $c$ is the speed of light,
$\beta\gamma = v/{c\sqrt{1-v^2/c^2}}$ is the 
relativistic correction for the time dilation.
We write $(\beta\gamma)_T=p_T/m_B$,
with $p_T$ the transverse momentum of the $B$ hadron and
$m_B$ the mass of the $B$ hadron.  The 
quantity $ct_{decay}$ is referred to as the proper decay
length.

The uncertainty in the measurement of the proper decay
length ($\sigma_{ct}$) has three terms:

\begin{equation}
\sigma_{ct} = ({m_B\over{p_T}})\sigma_{L_{xy}} \otimes 
ct({\sigma_{p_T}\over{p_T}}) \otimes ({L_{xy}\over{p_T}})\sigma_{m_B}
\end{equation}
where the $\otimes$ symbol indicates that the terms 
combine in quadrature.  The final term, which is 
proportional to the uncertainty on the $B$ hadron
mass $(\sigma_{m_B})$ is negligible in all cases.  
The first term is 
the uncertainty on the measured decay length.  This
depends upon the resolution of the detector as well
as the topology and momentum of the decay mode.  The
middle term, proportional to the uncertainty 
transverse momentum
of the $B$ hadron $(\sigma_{p_T})$ is effectively the
uncertainty in the time dilation correction.  For fully
reconstructed modes where all of the $B$ daughter particles
are accounted for, this term provides a negligible contribution
to the uncertainty on the proper decay time.  In the case
of partially reconstructed modes, where some fraction of
the $B$ daughters are not reconstructed, the uncertainty
on the $B$ hadron momentum becomes a significant contributor
to the lifetime uncertainty.

\begin{figure}
\center
\psfig{figure=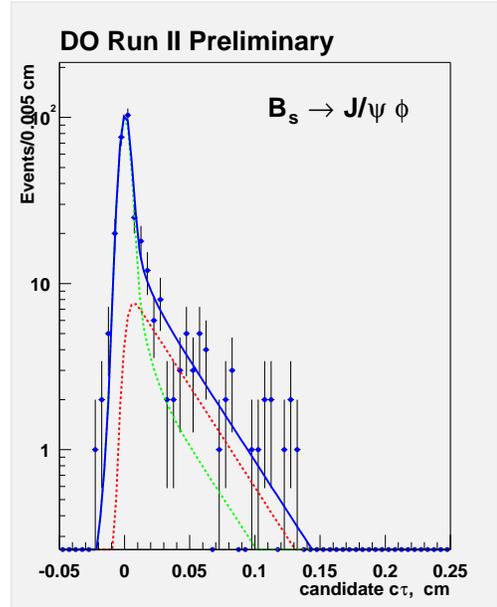,width=6.50truecm}
\caption{A measurement of the lifetime of the $B_s$ 
meson from the D\O\ experiment.  This result uses the
fully reconstructed $B_s \rightarrow J/\psi \phi $
mode, with $\phi\rightarrow K^+ K^-$.
The signal yield
is $133\pm 17$ events in $115\, \rm pb^{-1}$ of data.  The
solid curve is the combined signal + background fit to the
data, while the dashed curves are the data and background
components shown separately.  The background dominates at
zero decay time, but there is also a background contribution
coming from heavy flavor.}
\label{fig:d0taulamB}
\end{figure}

Fully reconstructed $J/\psi$
modes, such as $B^0\rightarrow J/\psi K^{*0}$, with 
$K^{*0}\rightarrow K^- \pi^+$ have the advantage of 
having small uncertainty in the $p_T$ of the $B$ hadron.
The drawback, however, is that the signal yields are
small due to the small branching ratio into the
color-suppressed $J/\psi$ mode.
Figure~\ref{fig:d0taulamB} shows a measurement
of the $B_s$ lifetime from D\O\ in the mode 
$B_s\rightarrow J/\psi \phi$, with $J/\psi\rightarrow
\mu^+ \mu^-$ and $\phi\rightarrow K^+K^-$.   
In fitting for the lifetime, it is necessary to account
for backgrounds from prompt sources as well as backgrounds
from heavy flavor sources.  In the case of the fully
reconstructed modes, the lifetime fit can additionally
utilize reconstructed mass information to properly weight
signal versus background events.  In the case of $J/\psi$
modes, the dominant backgrounds come from real $J/\psi$
decays from both prompt $c\overline{c}$ production as
well as $B$ decays.

\begin{figure}
\center
\psfig{figure=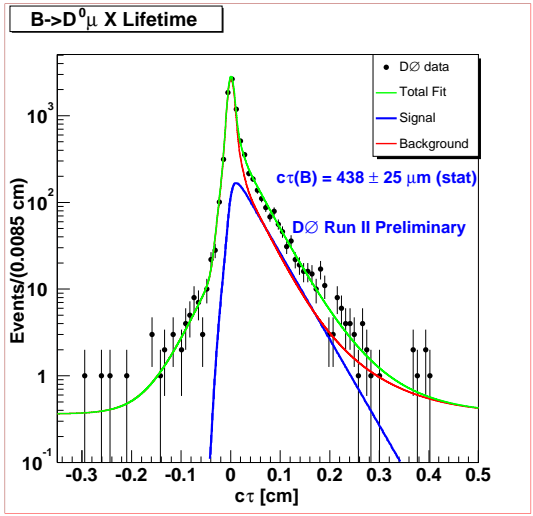,width=7.0truecm}
\caption{A measurement of the inclusive $B\rightarrow \mu D^0 X$
lifetime from the D\O\ experiment.  This inclusive lifetime has
contributions from all $B$ modes, although the sample is 
dominantly $B^+$ and
$B^0$ decays.}
\label{fig:tausemi}
\end{figure}

The statistical uncertainties on the D\O \ and CDF
lifetime  measurements
are not yet competitive with the current world average for
the $B_s$ and $\Lambda_b$ lifetimes.  With larger data
samples over the next 1-2 years, new results from the 
Tevatron will surpass the current level of precision.

Alternatively, semileptonic decays, such as
$B^0_s\rightarrow D^-_s\ell\nu_\ell$, with $D^-\rightarrow
\phi \pi^-$ provide larger signal yields but suffer
from uncertainty in the $B$ hadron $p_T$ due to the
unreconstructed neutrino.  With large data samples,
the semileptonic modes will begin to become systematics
limited due to the partial reconstruction, while
the statistics limited fully reconstructed modes 
will continue to provide improved sensitivity.

Figure~\ref{fig:tausemi} shows a measurement
of the inclusive $B$ lifetime in the $\mu D^0$ mode.
Since this is a partial reconstruction, backgrounds
can be more challenging than they are in the fully
reconstructed mode.  One technique to suppress backgrounds
is to demand the proper charge correlation between the
muon and the $D^0$ meson.  The charge of the charm
quark is carried through the $D^0$ decay ($D^0\rightarrow K^-\pi^+$
and $\overline{D^0}\rightarrow K^+\pi^-$) so the correlation
that the charge of the muon be the same as the charge of
the kaon can be enforced to reduce backgrounds.  This works
even without particle identification, because if the $K$ and
$\pi$ masses are assigned incorrectly they typically 
do not reconstruct
a $D^0$ mass.

One interesting background to the semileptonic
analysis is 
$c\overline{c}$ production where one charm hadron
decays semileptonically and the other fragments into
a $D^0$.  This background is only a problem if the
charm pair is produced at a very small opening angle, which
is exactly what occurs when a hard gluon splits into
a $c\overline{c}$ pair.  This ``gluon splitting'' 
contribution produces the ``right-sign'' charge 
correlation between the $D$ and muon.  Even though
the 
$D^0$ extrapolates back to the primary vertex in this
case, the ``fake'' $\mu D^0$ vertex can look like it
arose from a long-lived state.  Further studies of
$c\overline{c}$ production correlations are needed to
fully understand this background source.

\begin{figure}
\center
\psfig{figure=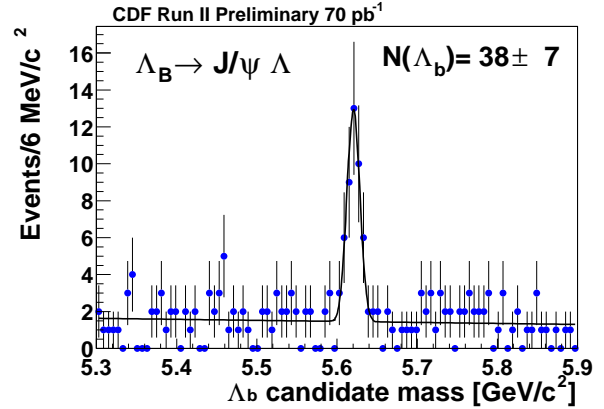,width=8.00truecm}
\caption{A measurement of the mass of the $\Lambda_b$ baryon
from the CDF experiment.  The fully reconstructed 
$\Lambda_b \rightarrow J/\psi \Lambda$, with $J/\psi \rightarrow \mu^+\mu^-$
and $\Lambda \rightarrow p\pi^-$ is shown here.  The mass scale
is calibrated with very high precision using $J/\psi$, $\psi(2S)$
and $\Upsilon \rightarrow \mu^+\mu^-$ decays.}
\label{fig:mass}
\end{figure}

\section{B Hadron Masses}

CDF has performed precision measurements of the masses of $B$
hadrons using fully reconstructed $B\rightarrow J/\psi X$ modes.
High statistics $J/\psi\rightarrow \mu^+\mu^-$ and $\psi(2S)\rightarrow
J/\psi \pi^+\pi^-$ are used to calibrate tracking momentum scale
and material in the tracking volume.  The results are tabulated
in Table~\ref{ta:bmasses}.  Figure~\ref{fig:mass} shows
the measurement of the $\Lambda_b$ baryon  mass.  Even with
relatively small statistics (less than 100 events in $B_s$
and $\Lambda_b$ modes)
these new results are the world's best measurements of
these masses.

\begin{table}
\caption{CDF results on masses of $B$ hadrons.  These results
come from fully reconstructed $J/\psi$ modes.  The first 
error is statistical, the second systematic.}
\label{ta:bmasses}
\begin{center}
\begin{tabular}{clc}
$B$ hadron  & Decay mode & Measured mass (${\rm MeV}/c^2$)   \\

\hline 
$B^+$ & $J/\psi K^+$ & $5279.32 \pm 0.68 \pm 0.94 $ \\
$B^0$ & $J/\psi K^{*0}$ & $5280.30 \pm 0.92 \pm 0.96 $ \\
$B^0_s$ & $J/\psi \phi$ & $5365.50 \pm 1.29 \pm 0.94 $ \\
$\Lambda_b$ & $J/\psi \Lambda $ & $5620.4\  \pm 1.6 \  \pm 1.2\  $ \\
\hline
\end{tabular}
\end{center}
\end{table}

\section{Hadronic Branching Ratios}

\subsection{Two-body Charmless $B$ Decays}

With the new SVT trigger, CDF has begun to measure $B$
decays with non-leptonic final states.  One set of modes of
particular interest are the charmless two-body modes.  Requiring
the final state to consist of two charged hadrons, the following
modes can be accessed at the Tevatron:
\begin{itemize} \itemsep=-6pt
\item $B^0\rightarrow \pi^+\pi^-$, $BR\sim 5\times 10^{-6}$ 
\item $B^0\rightarrow K^+\pi^-$, $BR\sim 2\times 10^{-5}$
\item $B_s\rightarrow K^+K^-$, $BR\sim 1\times 10^{-5}$
\item $B_s\rightarrow K^-\pi^+$, $BR\sim 2\times 10^{-6}$.
\end{itemize}
The $B^0$ states are accessible at the $e^+e^-$ facilities, 
but the $B_s$ modes are exclusive to the Tevatron.

The measurement presented here is the first observation of
the decay $B_s\rightarrow K^+K^-$.   As more data is accumulated,
the longer term goal from these modes is to search for direct
{\it CP}-violation as well as measure the $CKM$ angle 
$\gamma={\rm Arg}(V^*_{ub})$\rlap.\,\cite{fleisher}

\begin{figure}
\center
\psfig{figure=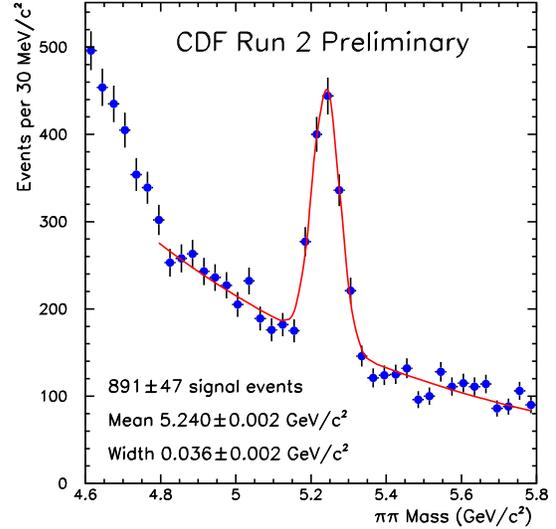,width=8.0truecm}
\caption{The reconstructed two-body $B\rightarrow h^+h^{-\prime}$ 
($h, h^\prime =K$ or $\pi$)
sample from CDF.  For this plot, all particles are reconstructed
as pions.  The peak consists of contributions from the four
modes listed in the text.  These pieces are extracted using
kinematic and particle identification information.  The
luminosity used for the figure shown here is ${\cal L}=190\, \rm
pb^{-1}$.  The results presented in the text are for a 
subset of this data.}
\label{fig:pipi}
\end{figure}

Figure~\ref{fig:pipi} shows the reconstructed signal where all
tracks are assumed to have the mass of the pion.   A clear peak is 
seen, and the width of the peak is significantly larger
($41\, {\rm MeV}/c^2$)
than the intrinsic resolution of the detector.  This additional
width is due to the $K^+\pi^-$ and $K^+K^-$ final states from $B^0$
and $B_s$ decays.  To extract the relative contributions, kinematic
information
and $dE/dx$ particle identification is used.\footnote{CDF has a 
time-of-flight system for particle identification with very
good $\pi$-$K$ separation for track $p_T < 1.6\, {\rm GeV}/c$.  The tracks
from these two-body decay modes have $p_T > 2 \, {\rm GeV}/c$ and 
therefore the time-of-flight system does not provide additional
particle identification information 
for this analysis.}  The particle identification
is calibrated from a large sample of $D^{*+}\rightarrow D^0\pi^+$ decays,
with $D^0\rightarrow K^-\pi^+$.  The charge of the pion from the $D^*$
uniquely identifies the kaon and pion, providing an excellent calibration
sample for the $dE/dx$ system.  Although the $K$-$\pi$ separation
is $1.3\sigma$, this is sufficient to extract the two-body $B$ decay
contributions.

The results from ${\cal L}= 65\, \rm pb^{-1}$ 
are shown in Table~\ref{ta:bpipi}.
The $B^0$ modes have been measured by CLEO, Babar and
Belle.  This is the first observation of the $B^0_s\rightarrow
K^+K^-$ decay mode.  Turning the yields into a ratio of branching
ratios, the result is:

\begin{displaymath}
{BR(B^0_s\rightarrow K^+K^-)\over{BR(B^0_d\rightarrow K^+\pi^-)}}
= 2.71 \pm 1.15 
\end{displaymath}
where the error is the combined statistical and systematic
uncertainty.  To calculate this ratio, information about the
relative production rates of $B^0_s$ and $B^0$ mesons
must be included.  The uncertainty includes the uncertainty
on the relative production fractions.

\begin{table}
\caption{CDF results on two-body charmless $B$ decays.
The yields reported in this table are extracted by fitting
the decay information for the events, including kinematic
and particle identification information.  Yields shown here
are for ${\cal L} \sim \! 65 \, \rm pb^{-1}$.}
\label{ta:bpipi}
\begin{center}
\begin{tabular}{lr}
Mode & Fitted yield (events) \ \ \ \ \   \\

\hline 
$B^0\rightarrow K^-\pi^+$ & $148\pm 17(stat.)\pm 17(syst.)$\\
$B^0\rightarrow \pi^+\pi^-$ & $39\pm 14(stat.)\pm 17(syst.)$\\
$B^0_s\rightarrow K^+K^-$ & $90\pm 17(stat.)\pm 17(syst.)$\\
$B^0_s\rightarrow K^-\pi^+$ & $3\pm 11(stat.)\pm 17(syst.)$\\
\hline
\end{tabular}
\end{center}
\end{table}

\subsection{$\Lambda_b \rightarrow \Lambda_c \pi^-$ Branching Ratio}

\begin{figure}
\center
\psfig{figure=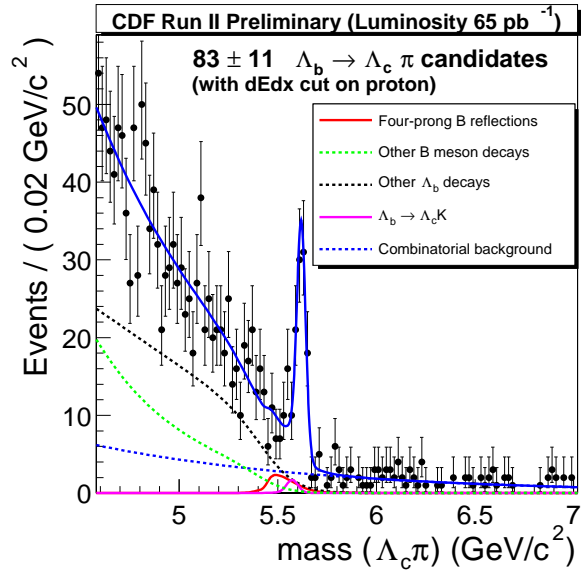,width=8.0truecm}
\caption{The CDF fully reconstructed $\Lambda_b\rightarrow \Lambda_c \pi$,
with $\Lambda_c \rightarrow pK^-\pi^+$.
The features in the sidebands are discussed in the text.  For
this plot, an additional particle identification cut on the
proton was made that was not used in the extraction of the 
branching ratio.}
\label{fig:lblcpi}
\end{figure}

Using the SVT trigger, CDF has begun to measure
$b$-baryon states.  Figure~\ref{fig:lblcpi} shows a
clean signal of the decay $\Lambda_b \rightarrow \Lambda_c \pi^-$,
with $\Lambda_c \rightarrow pK^-\pi^+$.  The reconstructed
invariant mass plot has a very interesting structure, with
almost no background above the peak and a background that
rises steeply in going to lower mass.  This structure is 
somewhat unique to baryon modes, which are the most massive
weakly decaying $B$ hadron states.  Because the SVT trigger
specifically selects long-lived states, most of the backgrounds
are coming from other heavy flavor ($b$ and $c$) decays.
Since there are no weakly decaying 
$B$ hadrons more massive than the $\Lambda_b$,
there is very little background above the peak.  On the other
hand, going to masses below the peak, lighter $B$ mesons begin to 
contribute.  The background in this mode is growing at lower
masses because there is more phase space for $B^+$, $B^0$,
and  $B^0_s$ to contribute.

To extract the number of signal events, $b\overline{b}$
Monte Carlo templates are used to account for the 
reflections seen in the signal window.  The shapes of 
these templates are fixed by the simulation, but their
normalization is allowed to float.
The number of fitted signal events in this analysis is
$96 \pm 13(stat.)^{+6}_{-7} (syst.)$.  The primary result
from this analysis is a measurement of the $\Lambda_b
\rightarrow \Lambda_c \pi^-$ branching ratio relative to
the $B^0\rightarrow D^-\pi^+$ mode.  We can take that 
ratio, along with PDG 2002\cite{pdg} values for 
measured branching ratios and production fractions, and 
extract 
the  
branching ratio
\begin{displaymath}
BR(\Lambda_b\rightarrow \Lambda_c \pi^-) =  
(6.5\pm 1.1 \pm 0.9  \pm 2.3)\times 10^{-3},
\end{displaymath}
where the errors listed are statistical, systematic and the
final uncertainty is arising from the 
uncertainty in the $B^0\rightarrow D^-\pi^+$ branching
ratio.

\subsection{Observation of the $X(3872)$ State}

At this Lepton-Photon Symposium, the Belle collaboration announced
the observation of a neutral state decaying into
$J/\psi \pi^+\pi^-$ with a mass of 3872 MeV\rlap.\,\cite{bellex3872}
This state may be the  $1^3D_2$  $c\overline{c}$ bound state,
although the observed mass is higher than expected 
for that state.  It has also been hypothesized that this
is a loosely bound $D\overline{D}^*$ bound state, since
the mass is right at the $D\overline{D}^*$ threshold.

Belle observes this state in $B$ decays. Their 
observation further indicates that the state is narrow,
and favors large $\pi^+\pi^-$
mass in the decay.

\begin{figure}
\center
\psfig{figure=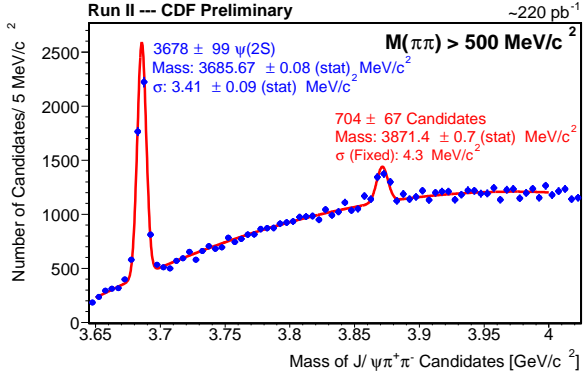,width=8.50truecm}
\caption{The CDF observation of a neutral state of mass
$3871\, {\rm MeV}/c^2$ decaying to $J/\psi \pi^+\pi^-$  This
is a confirmation of the state first reported by the
Belle collaboration at this conference.  The distribution
shown here includes a requirement that the dipion mass
have $M(\pi \pi)>500 \, {\rm MeV}/c^2$.  The signal is clear
and significant without any cut on the dipion mass.}
\label{fig:x3872}
\end{figure}

CDF has searched for this state using a sample of
${\cal L}\sim\! 220 \, \rm pb^{-1}$ and sees a
clear signal with a mass of 
\begin{displaymath}
3871.4 \pm 0.7(stat.) \pm 0.4 (syst.) {\rm MeV}/c^2.
\end{displaymath}
Figure~\ref{fig:x3872} shows the CDF $J/\psi \pi^+\pi^-$
mass spectrum.  This plot originates from a parent sample
of approximately $2.2$ million  $J/\psi \rightarrow \mu^+\mu^-$ 
decays.  The large peak at $3.685\, {\rm MeV}/c^2$ 
is the $\psi(2S)$\rlap.\,\cite{cdfx3872}

The plot shows events where the dipion mass was required
to be greater than $500 \, {\rm MeV}/c^2$.  The significance
of the signal without this cut is greater than $10\sigma$.
CDF also reports that the state is narrow.  The observed
width of $4.3\, {\rm MeV}/c^2$ is consistent with detector
resolution.  Further studies are underway to investigate the
$\pi^+\pi^-$ mass distribution as well as to determine 
whether or not the CDF signal is coming from a prompt 
source or though $B$ decays.  Since all angular momentum
states are accessible in high energy $p\overline{p}$ collisions,
it is possible that this state is directly produced at the
Tevatron, while it cannot be directly produced in $e^+e^-$
collisions.

\section{Mixing}

In the $K^0$ and $B^0$ systems, particle-antiparticle mixing has been observed and measured with
great precision.  This mixing is understood to occur
because the weak interaction eigenstates are not the
same as the strong interaction (or flavor) eigenstates.
The weak eigenstates are then linear combinations of
the flavor eigenstates, giving rise to an oscillation
frequency that is proportional to the mass difference
between the heavy and light states.


Recently, the Babar and Belle 
experiments have made very precise 
measurements of $B^0\overline{B^0}$ mixing
and have significantly improved the world 
average.  The mixing parameter is typically reported 
in terms of the heavy/light mass difference.  For the
$B^0$ system, the world average is:\cite{hfag}
\begin{displaymath}
\Delta m_d = 0.502 \pm 0.006 \, \rm ps^{-1}.
\end{displaymath}
From this, we see that a beam of pure $B^0$
mesons would result in a beam of pure $\overline{B^0}$
mesons in time $\Delta m_d t_{mix} = \pi$, which indicates
that $t_{mix}\sim\! 4.1$ lifetimes, indicating
that the oscillation is rather slow.

\begin{figure}
\center
\psfig{figure=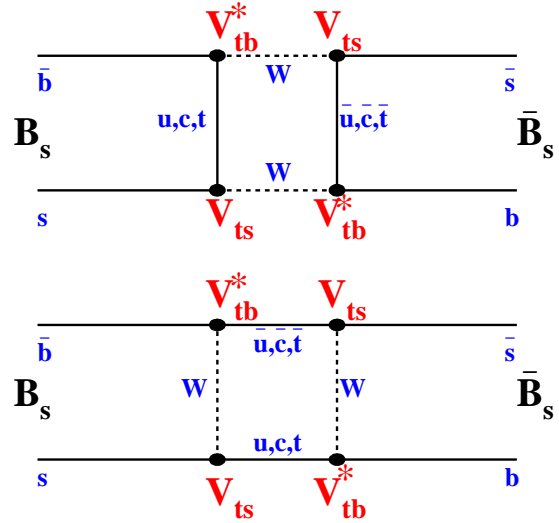,width=7.50truecm}
\caption{The box diagrams for the $B^0_s\rightarrow \overline{B^0_s}$
transition.  All up-type quarks ({\it u,c,t}) are included in the box,
but since these terms are proportional to the mass of the quark
in the box, the top quark dominates.  Therefore the $V_{ts}$ 
element is relevant for $B_s$ mixing and the $V_{td}$ element is
relevant for $B^0$ mixing.}
\label{fig:mix}
\end{figure}

Mixing proceeds via a second-order weak transition as shown
in Fig.~\ref{fig:mix}.  The box diagram includes the
$V_{td}$ matrix element for $B^0/\overline{B^0}$ mixing,
which is replaced by $V_{ts}$ for $B^0_s/
\overline{B^0_s}$ mixing.  Experimentally, we know that
$V_{ts}$ is larger than $V_{td}$:
\begin{displaymath}
Re(V_{ts})\simeq 0.040 > Re(V_{td})\simeq 0.007,
\end{displaymath}
so we expect the $B^0_s$ system to oscillate with a 
much higher frequency than the $B^0$ system.
Indeed this is the case, the $B^0_s$ system oscillates
so quickly that the oscillations have not yet been 
resolved.  The current combined world limit is:\cite{hfag}
\begin{displaymath}
\Delta m_s > 14.4 \, {\rm ps^{-1}}~@ 95\%~{\rm CL},
\end{displaymath}
which means a beam of $B^0_s$ mesons would fully become
a beam of $\overline{B^0_s}$ mesons in less than $1/7^{\rm th}$
of one lifetime!
For the next several years, the Tevatron will be the
exclusive laboratory for $B_s$ meson studies, including
the search for $B_s$ mixing. 

\subsection{Measuring Mixing}

To measure $B_s$ mixing, four ingredients are needed.
\begin{itemize}\itemsep=-3pt
\item {\bf  Flavor at the time of production.}  It is
necessary to know whether the meson was produced as a
$B^0_s$ or a $\overline{B^0_s}$.
\item {\bf Flavor at the time of decay.}  It is also
necessary to know whether the meson was a $B^0_s$ or
$\overline{B^0_s}$ when it decayed.  This, combined with
the flavor at time of production, tells us whether the
$B_s$ had decayed as mixed or unmixed.\footnote{A meson
that ``decayed as unmixed'' could have mixed and mixed
back, undergoing one or more complete cycles.  This necessarily 
comes out of the time-dependent analysis.}
\item {\bf Proper decay time.}  It is necessary to know
the proper decay time for the $B_s$, since we are 
attempting to measure the  probability to mix as a function
of decay time.  The $B_s$ system mixes too quickly to 
resolve using time-integrated techniques.
\item {\bf Large $B_s$ samples.}  We must map
out the probability to mix as a function of decay time 
for at least part of the decay time spectrum.  Because each 
the previous three items have shortcomings requiring
more statistics, 
this analysis requires large samples of $B_s$ decays.
\end{itemize}

In the following subsections, we will discuss each of
these pieces necessary to measure $B_s$ mixing.

\subsection{Flavor Tagging}

The first two items in our list of requirements have to
do with determining the flavor of the $B_s$ meson at
the time of production and at the time of decay, referred  
to as initial-state and final-state flavor
tagging.  

For initial-state flavor tagging, we 
infer the flavor of the $B_s$ meson at the time of 
production from other information in the event.  Here
we can take advantage of what we know about $b\overline{b}$
production.  By measuring the flavor of the other $B$ hadron
in the event, we can infer the flavor of the $B_s$ at the
time of production.  This technique is imprecise, and also 
suffers from the fact that quite often ($\sim\! 75\%$) the
other $B$ hadron is outside the acceptance of the detector.

\begin{figure}
\center
\psfig{figure=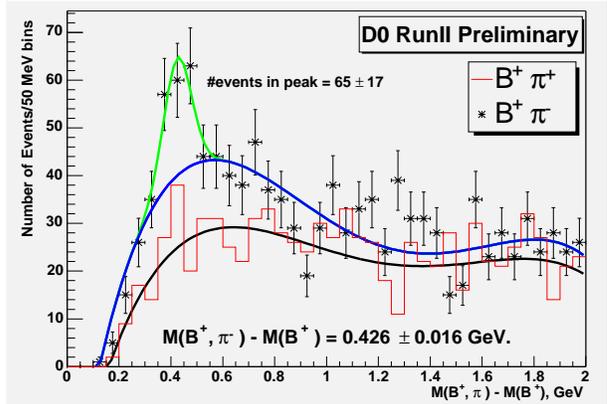,width=8.0truecm}
\caption{D\O\ same-side tagging plot.}
\label{fig:bstar}
\end{figure}

Another technique is to look at fragmentation tracks near
the $B_s$ meson.  For a $\overline{b}$ quark to become a
$B^0_s$ meson, it must grab an $s$ quark from the vacuum.
When the $s$ is popped from the vacuum, an $\overline{s}$ is
popped with it, which could potentially turn into a $K^+$
meson.  We can then use the charge of the kaon to infer the
flavor of the $B$ hadron.  Again, this is an inexact 
technique, since other fragmentation tracks can confuse 
this correlation
and also the charge information could be lost into neutral
particles, like $K^0_S$.  Figure~\ref{fig:bstar} shows an
example of same-side tagging.  In this case, a fully
reconstructed $B^+\rightarrow J/\psi K^+$ sample is used,
where the flavor of the $B$ hadron is known.  The plot
then shows the mass difference between the $\pi$-$B^+$
and the $B^+$.  A clear opposite-sign excess is seen over
the entire range of the plot, which is attributed to the
fragmentation correlation.  In addition, a clear peak 
is seen near $0.4\, {\rm GeV}/c^2$ which is attributed to
the $B^{**}$ state.

The efficacy of the initial-state flavor tagging is classified
by the tagging power: $\epsilon D^2$, where $\epsilon$ is the fraction
of times the algorithm was able to arrive at a tagging decision and 
$D$ is the dilution, which is a measure of the probability that
the tag is correct.  The dilution is written as: $D=(N_R-N_W)/(N_R+N_W)$,
where $N_R$($N_W$) are the number of right (wrong) tags.  The 
dilution is related to the mistag fraction, $w$ as $D=1-2w$.
Tagging power is proportional to $D^2$ because 
incorrectly measuring the sign  removes the event from
the ``correct'' charge bin and puts the
event into the ``wrong'' charge bin.

For Babar and Belle, $\epsilon D^2\simeq 27\%$,
whereas for the Tevatron $\epsilon D^2 \simeq 5\%$.  The
large difference  arises because of the nature and cleanliness of
the $e^+ e^-$ environment.  As an example, if an experiment has
1000 signal events with $\epsilon D^2 = 5\%$, then the statistical
power of that sample is equivalent to a sample of 50 events
where the tag is known absolutely.  

For final-state flavor tagging, there are three primary classes of $B_s$
decays:
\begin{itemize}\itemsep=-6pt
\item $B^0_s\rightarrow D^-_s \pi^+$, with $D^-_s\rightarrow \phi \pi^-$,
\item $B^0_s\rightarrow D^-_s \ell^+ \nu_\ell$, with $D^-_s\rightarrow \phi \pi^-$,
\item $B^0_s\rightarrow J/\psi \phi$, with $J/\psi \rightarrow \mu^+\mu^-$.
\end{itemize}
In the first two cases, the flavor of the $B_s$ is immediately evident
from the charge of the decay products, which are referred to as
a ``self-tagging'' final states.  In the third case, there is no
way know whether the meson decayed as a $B^0_s$ or $\overline{B^0_s}$,
so charge-symmetric modes are of no use for the mixing 
analysis.\footnote{The charge-neutral $B_s$ modes are important for
other analyses, such as the search for a lifetime difference $\Delta \Gamma_s$
in {\it CP}-even and {\it CP}-odd decays.}

\subsection{Proper Decay Time Resolution}

We know the $B_s$ oscillates very quickly, therefore we need
proper time resolution that is smaller than the oscillation
frequency.  As discussed in detail in Sec.~\ref{se:lifetimes},
the two primary components contributing to the proper time
resolution are vertex ($L_{xy}$) resolution
and the  time dilation correction ($(\beta \gamma)_T= p_T/m_B$).

For the semileptonic samples, the time dilation correction
factor limits the proper time resolution.  For fully 
reconstructed samples (no missing neutrino) the time
dilation correction is a negligible effect and only the
$L_{xy}$ resolution contributes to the uncertainty on
the proper decay time measurement. 

If the true value of $\Delta m_s$ is close to the current 
limit $\Delta m_s \! \sim \! $14$-$18$ \, \rm ps^{-1}$, then both
fully reconstructed and semileptonic samples will contribute 
to the measurement of $\Delta m_s$.  However,
if the true value of $\Delta m_s$ is $20\, \rm ps^{-1}$ or
higher, then the proper time resolution  becomes
the limiting factor in resolving the oscillations.  At 
$\Delta m_s$ values this high, only the fully reconstructed
samples are useful, and in fact the vertex resolution 
becomes the limiting factor.

\subsection{Yields}

\begin{figure}
\center
\psfig{figure=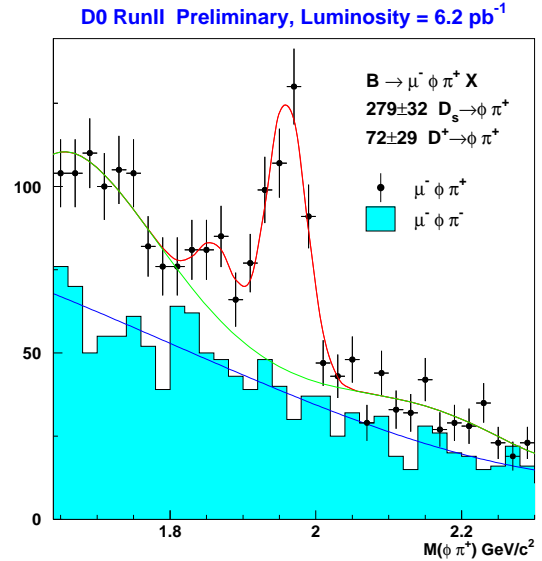,width=8.0truecm}
\caption{Semileptonic $B^0_s$ decays can be partially
reconstructed in the mode $B^0_s \rightarrow \ell^+ D^-_s X$,
with $D^-_s\rightarrow \phi \pi^-$ and $\phi\rightarrow K^+K^-$.
The distribution shown here is a sample of $\mu \phi \pi$ events
from D\O.  The two peaks correspond to the $D^+$ and $D^+_s$ states,
which can both decay to $\phi \pi^+$.  This mode will provide a
large sample of semileptonic $B^0_s$ decays to search for 
$B^0_s/\overline{B^0_s}$ mixing.}
\label{fig:lds}
\end{figure}

As stated previously, the statistics of the semileptonic
samples are higher than those of the fully reconstructed 
modes.   For lower values of $\Delta m_s$ the larger 
semileptonic event yields somewhat offset the poorer proper
time resolution.  As a consequence, both CDF and D\O \ are 
continuing to acquire semileptonic samples, both for
flavor tagging calibration and for the $B_s$ mixing search.
A semi-muonic sample enriched in $B_s$ decays is shown
in Fig.~\ref{fig:lds}, demonstrating that large semileptonic
samples are being acquired.

\begin{figure*}[t]
\psfig{figure=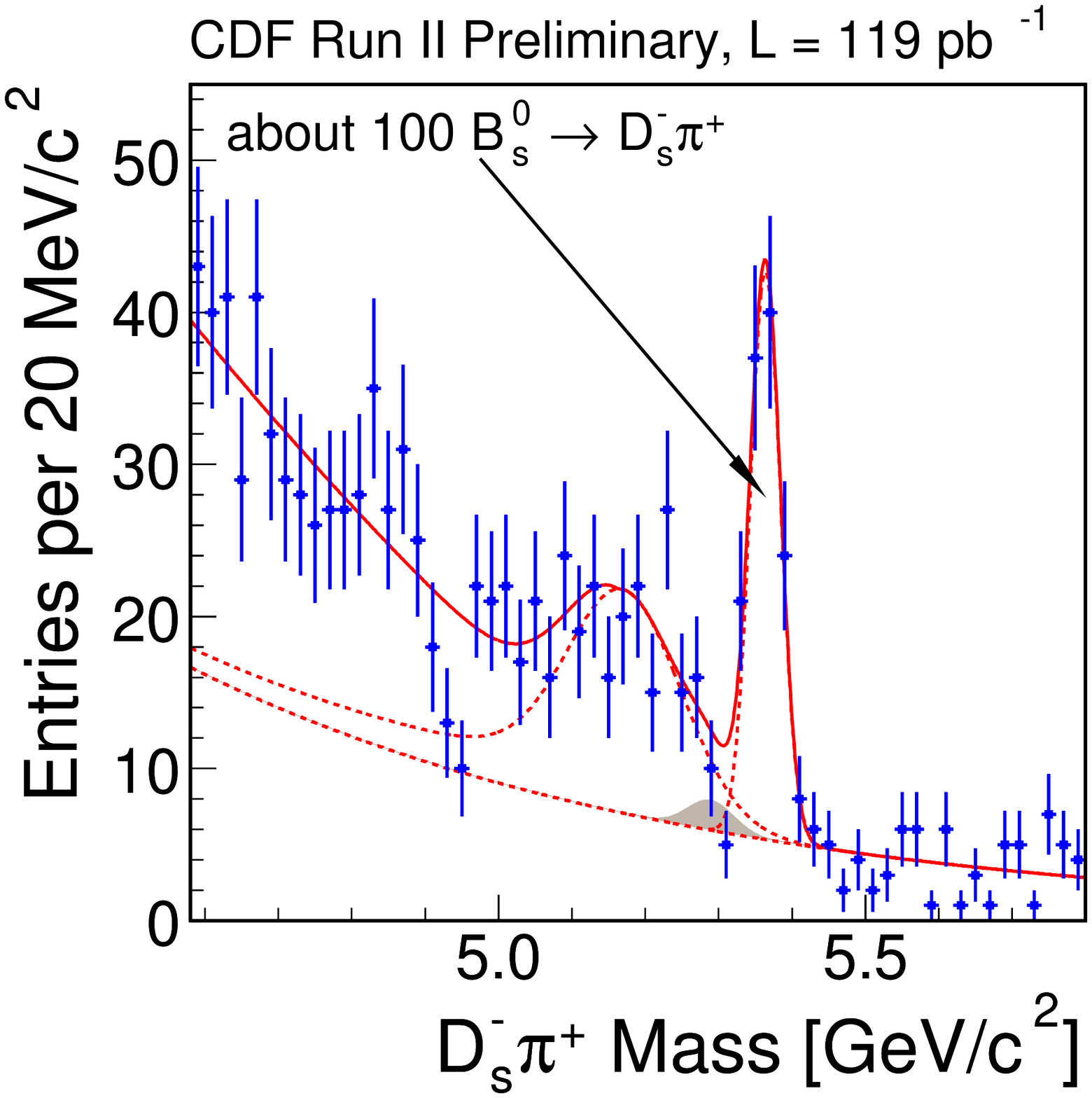,width=3.1truein}
\hspace*{0.1truein}
\psfig{figure=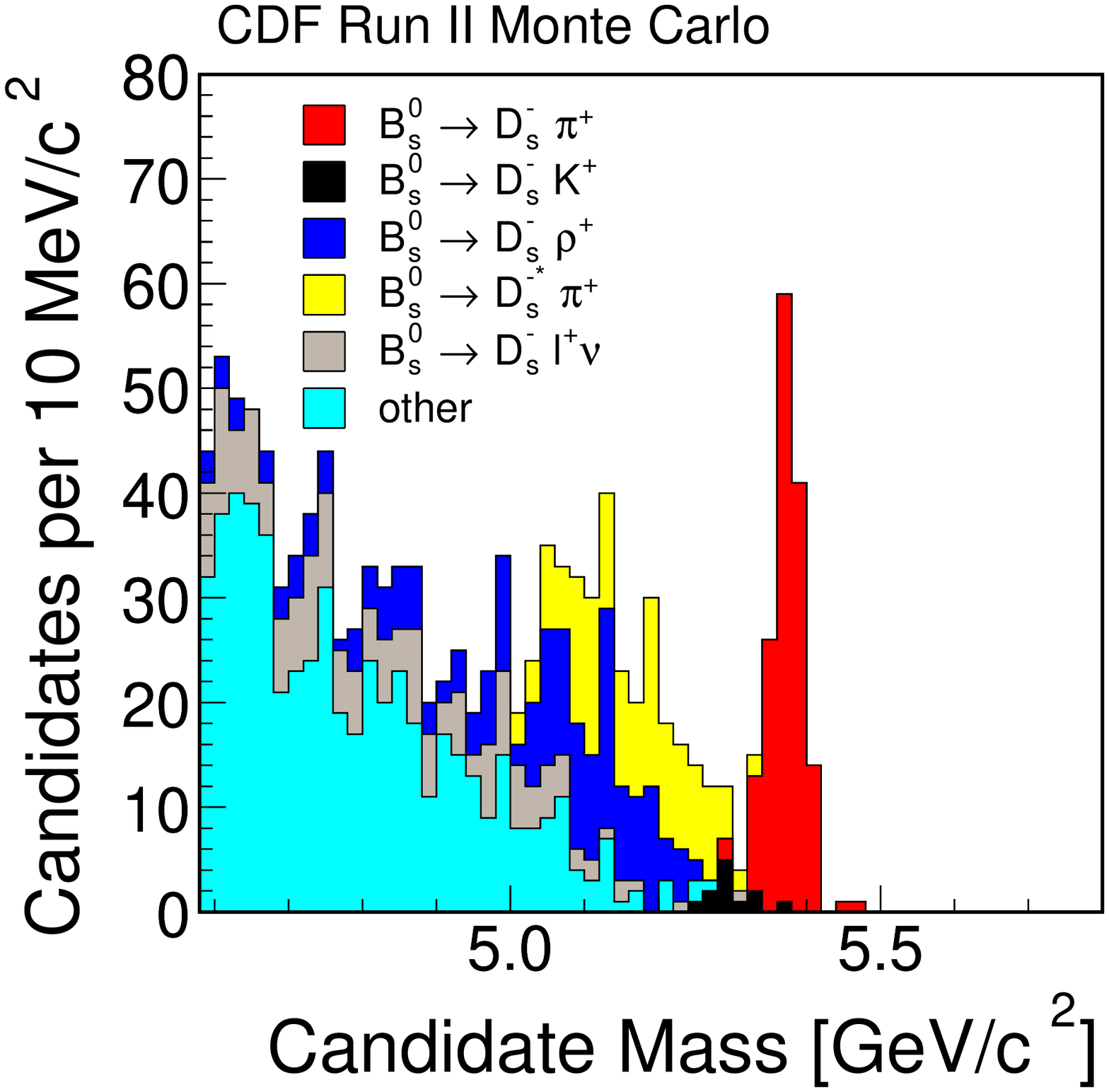,width=3.1truein}
\caption{Fully reconstructed $B^0_s$ decays from CDF using
the SVT trigger.  The decay chain is $B^0_s\rightarrow
D^-_s \pi^+$, with $D^-_s \rightarrow \phi \pi^-$ and 
$\phi \rightarrow K^+K^-$.  For the mixing analysis, the
proper time resolution will be better in this mode, because
the error on the time-dilation factor is negligible.  The
challenge is the limited statistics.  CDF is currently trying
to reconstruct other modes to supplement this sample.
\label{fig:bsdspi} \vspace*{6pt}}
\end{figure*}

The fully reconstructed states offer fewer signal events, but
the improved proper time resolution compensates for this 
at values of $\Delta m_s$ that we are considering.  CDF
has accumulated a sample of fully reconstructed $B_s$ decays
as shown in Figure~\ref{fig:bsdspi}. The plot on the left
is the signal, a clear $B_s$ peak can be seen.  The broad
peak below the $B_s$ is the $B_s\rightarrow D^*_s \pi$, where
the photon from the $D^*_s$ decay is not reconstructed.  
The plot on the right shows the expected contributions from
a $b\overline{b}$ Monte Carlo.  In the data, the 
signal and sidebands are
fit using the shapes from the Monte Carlo letting the normalizations
float.  The Monte Carlo clearly provides a very good description
of the signal and heavy flavor backgrounds.  This is possible
because the SVT trigger provides very pure heavy flavor 
samples.  Even though the hadron collider environment is 
challenging, clean samples of heavy flavor decays can be isolated.

\subsection{$B_s$ Mixing Status and Prospects}

Both experiments have now commissioned the detectors
and accumulated the first portion of Tevtron Run~2 data.
If we take the current performance of the trigger, 
reconstruction and flavor tagging, it appears that with
a sample of ${\cal L}\sim\! 500 \, \rm pb^{-1}$, the
$B_s$ sensitivity will be comparable to the current 
combined world limit.   To observe or exclude a value
of $\Delta m_s > 15 \, \rm ps^{-1}$ will require additional
data and improvements along with further progress on triggering,
reconstruction and flavor tagging.

With modest improvements to the current running configuration,
we estimate that it will take $2$-$3\, \rm fb^{-1}$ of 
integrated luminosity to ``cover'' the region of $\Delta m_s$
that is currently preferred by indirect fits\rlap.\,\cite{indirect}
It is important to recall that the current combined world 
limit consists of contributions from 13 different measurements,
and is the culmination of the LEP, SLD and CDF Run I programs.
Thanks to additional luminosity and upgraded detectors, it will
be possible to extend this search for $B_s$ mixing, however this
is a very challenging measurement that will take time, effort and
a significant data sample.


\section{Summary}

The results shown in this summary provide a snapshot of
the heavy flavor results coming from the Tevatron.  The 
period of commissioning is now complete, and the experiments
are slightly more than one year into a multi-year run 
that will continue to accumulate large data samples.
The upgraded detectors are performing well, and many of
the upgrades specific to $B$ physics are beginning to 
pay off.  Over the next several years, CDF and D\O\ will
make significant contributions in our understanding of
production and decay of heavy flavors.

\section*{Acknowledgments}
I would like to thank the Lepton-Photon 2003 organizers 
for the opportunity to speak at this excellent  conference.
I would also like to thank and acknowledge the collaborators
of the Babar, Belle, CDF and D\O \ experiments.  This
work is supported by the U.S. Department of Energy Grant
DE-FG02-91ER40677.

\clearpage
\twocolumn[
\section*{DISCUSSION}
]

\begin{description}
\item[ Jonathan L. Rosner] (University of Chicago):
What are the prospects for seeing fully reconstructed $B_c$
mesons, e.g. in $J/\psi \pi^\pm$?

\item[Kevin Pitts:]
In Run I with ${\cal L} \simeq 100\, \rm  pb^{-1}$, the $B_c$ was observed through
the decay $B_c \rightarrow J/\psi \ell \nu_\ell$, with $\ell=e,\mu$ and 
$J/\psi \rightarrow
\mu^+\mu^-$, providing a tri-lepton final state.  In addition, 
there was a hint of a signal in $B_c \rightarrow J/\psi \pi^\pm$.
In both of these modes, the $b$ quark decays to charm, 
providing the $J/\psi$
in the final state.
In Run II,  with ${\cal L}\sim \! 220 \, \rm pb^{-1}$ 
already on tape the prospects are quite good
both for semileptonic decay and for fully reconstructed
decays.
In addition, with the enough statistics in the 
hadronic trigger, it is likely that
CDF can reconstruct the $B_c$ in modes where the charm
quark decays, such as $B^+_c \rightarrow B^0_s \pi^+$.  This
mode might be especially interesting as a new tagging
mode for $B^0_s$ mixing.

\item[Vivek Sharma] (University of California at San Diego):
What is the timescale for a $10\%$ measurement of $\Lambda_b$ lifetime?
In particular can you use the various fully reconstructed hadronic
$\Lambda_b$ modes given the reflection and impact parameter bias?

\item[Kevin Pitts:]
Performing a simple average of the CDF and D\O\ measurements of the
$\Lambda_b$ lifetime in $\Lambda_b\rightarrow J/\psi \Lambda$, the
result is $\tau_{\Lambda_b} = 1.13 \pm 0.18 \, \rm ps$, which is
a $16\%$ measurement on the $\Lambda_b$ lifetime.  These results
do not yet use the entire Run~II data samples available.  
Assuming all errors 
scale like $1/\sqrt{N}$, the combined result from the two experiments
using the full ${\cal L} \sim \! 220 \, \rm pb^{-1}$ on tape as
of this conference, the combined result would be a $10\%$ measurement
on the $\Lambda_b$ lifetime.  With more data coming in, it seems
safe to expect that each experiment will have a measurement in the
neighborhood of $10\%$ by the summer of 2004.

As for the hadronic modes, significant progress has been made 
in understanding both the reflections and the impact parameter
bias.   The understanding of the reflections has been shown in
the context of the $\Lambda_b \rightarrow \Lambda_c \pi$ branching
ratio measurement.  Understanding the lifetime bias coming from
the SVT trigger is a necessity for $B_s$ mixing, and all lifetime
measurements will benefit from ongoing progress on that front.

\item[Vera Luth] (SLAC):
For charm or $B$ decays you can normalize your measured 
$BR$
to other measurements. How are you planning to obtain
absolute $BR$ and production rates for $B_s$, $\Lambda_b$, etc?

\item[Kevin Pitts:]
We can still normalize our branching ratios to other modes,
but you point out an additional complication coming about 
due to the lack of precision in our measurements of the relative
production fractions.  For example, in the CDF measurement of 
the branching ratio for $\Lambda_b \rightarrow \Lambda_c \pi$,
the number of signal events in two modes is measured:
$N(\Lambda_b \rightarrow \Lambda_c \pi)$ and
$N(B^0 \rightarrow D^-\pi^+)$.  The efficiencies and
acceptances are calculated, and the PDG values for
the $BR(D^-\rightarrow K^-\pi^+\pi^+)$ and
$BR(\Lambda_c \rightarrow pK^- \pi^+$ are used.  Combining this,
the measured ratio of corrected  yields gives:
\begin{displaymath}
{f_{baryon} \times BR(\Lambda_b \rightarrow \Lambda_c \pi^-)\over{
f_d \times BR(B^0\rightarrow D^- \pi^+)}}
\end{displaymath}
where $f_{baryon}$ ($f_d$) are the fraction of produced $\Lambda_b$ ($B^0$)
hadrons in $p\overline{p}$ collisions. 

The extra piece that comes about in taking this ratio is
the ratio of 
production fractions. These are not known very well, and it is
important to improve upon our knowledge of the production fractions.
This is easier said than done.  One way to improve upon our knowledge
of $f_s$/$f_d$ is to measure $\overline{\chi}$, the time-integrated
$B/\overline{B}$ mixing parameter.  Other analyses have looked at
lepton+$D_s$, lepton+$D^0/D^\pm$ and lepton+$\Lambda_c$ 
correlations to attempt to
extract the species fractions.   Once statistics have improved, the
uncertainty in species fractions will be one of the dominant uncertainties
in attempting to extract absolute branching ratios.

\end{description}


\begin{thebibliography}{99}


\bibitem{d0}  D\O\  Collaboration, FERMILAB-Pub-96/357-E, 1996.

\bibitem{cdf} CDF Collaboration, 
FERMILAB-Pub-96/390-E, 1996;
Fermilab-Proposal-909, 1998.

\bibitem{bxsec}CDF Collaboration (D. Acosta {\it et al.}),  {\em Phys. Rev.} {\bf D65}, 052005, (2002).

\bibitem{charmxsec} CDF Collaboration (D. Acosta \it et al.\rm),
FERMILAB-PUB-03-217-E, hep-ex/0307080, July 2003.

\bibitem{nlo} M. Cacciari and P. Nason, hep-ph/0306212, {\em JHEP} 0309:006 (2003), June 2003.

\bibitem{taubabar}BABAR Collaboration (B. Aubert \it et al.\rm), {\em Phys. Rev. Lett.} {\bf87}, 201803 (2001);  BELLE Collaboration (K. Abe \it et al.\rm), {\em Phys. Rev. Lett.} {\bf 88}, 171801 (2002).

\bibitem{hfag}Heavy Flavor Averaging Group, \hfill http://www.slac.\\stanford.edu/xorg/hfag/index.html.


\bibitem{fleisher} R. Fleischer, {\em Phys. Lett.} {\bf B459}, 306-320 (1999).

\bibitem{pdg}Particle Data Group, http://pdg.lbl.gov/.

\bibitem{bellex3872} BELLE Collaboration (K. Abe \it et al.\rm), hep-ex/0308029, August 2003; S.-K. Choi \it et al.\rm, hep-ex/0309032, September 2003, submitted to PRL.

\bibitem{cdfx3872}CDF Collaboration, \hfill http://www-cdf.fnal.gov/\\physics/new/bottom/030224.blessed-x3872/

\bibitem{indirect} M. Ciuchini, \it et al.\rm, hep-ph/0307195, April 2003.

\end{thebibliography}
\end{document}